\documentclass[aps,prd,showpacs,preprintnumbers,amssymb,amsmath]{revtex4}
\usepackage{epsfig,bm,color,subfigure} 
\def \beq{\begin{equation}}
\def \eeq{\end{equation}}
\def \beqa{\begin{eqnarray}}
\def \eeqa{\end{eqnarray}}
\def \c {\chi}
\def \etc{{\sl etc.\/}}
\def \eg{{\sl e.g.\/}}
\def \ie{{\sl i.e.\/}}
\def \eg{{\sl e.g.\/}}
\def \etal{{\sl et al.\/}}
\def \jhep{{\sl J.\ H.\ E.\ P.\/}}
\def \np{{\sl Nucl.\ Phys.\/}}
\def \pl{{\sl Phys.\ Lett.\/}}
\def \pr{{\sl Phys.\ Rev.\/}}
\def \prl{{\sl Phys.\ Rev.\ Lett.\/}}
\begin{document}
\title{Robustness of baryon-strangeness correlation and related ratios
of susceptibilities}
\author{Swagato \surname{Mukherjee}}
\email{swagato@tifr.res.in}
\affiliation{Department of Theoretical Physics, Tata Institute of Fundamental
             Research,\\ Homi Bhabha Road, Mumbai 400005, India.}
\begin{abstract}
Using quenched lattice QCD simulations we investigate the continuum
limit of baryon-strangeness correlation and other related conserved
charge-flavour correlations for temperatures $T_c<T\le2T_c$. By working
with lattices having large temporal extents ($N_\tau=12, 10, 8, 4$) we
find that these quantities are almost independent of the lattice
spacing, \ie, robust. We also find that these quantities have very mild
dependence on the sea quark mass and acquire values which are very close
to their respective ideal gas limits. Our results also confirm robustness
of the Wroblewski parameter. 
\end{abstract}
\pacs{12.38.Aw, 11.15.Ha, 05.70.Fh}
\preprint{TIFR/TH/06-16}
\maketitle
\section{Introduction} \label{sc.intro}
The recent results from the Relativistic Heavy Ion Collider (RHIC)
\cite{rhic-results} indicate the formation of a thermalized medium
endowed with large collective flow and very low viscosity \cite{teaney}.
These findings suggest that Quark Gluon Plasma (QGP) is a strongly
interacting system for temperatures close to its transition temperature
($T_c$). Apart from the experimental indications, the most convincing
evidence in favour of the existence of a strongly interacting QGP comes
form the lattice QCD simulations \cite{lattice-results, swagato}. These
non-perturbative studies show that the thermodynamic quantities, like
pressure and energy density, deviate form there respective ideal gas (of
free quarks and gluons) values by about $20\%$ even at temperature
$T=3T_c$. On the other hand, other lattice studies indicate the
smallness of the viscous forces in QGP \cite{nakamura}. All these
results point to the fact that close to $T_c$ nature of QGP is far from
a gas of free quarks and gluons.

In order to uncover the nature of QGP in the vicinity of $T_c$ and also
to understand the underlying physics of these lattice results many
different suggestions have been made over the last decade. Descriptions
in terms of various quasi-particles \cite{quasi-results,bluhm}, resummed
perturbation theories \cite{pert-results}, effective models
\cite{effect-results} \etc\ are few among many such attempts. Apart
from all these, the newly proposed model of Shuryak and Zahed
\cite{shuryak} has generated considerable amount of interest in the
recent years. Motivated by the lattice results for the existence of
charmonium in QGP \cite{charmonium}, this model proposed a strongly
interacting chromodynamic system of quasi-particles (with large thermal
masses) of quarks, anti-quarks and gluons along with their numerous
bound states. As different conserved charges, \eg, baryon number ($B$),
electric charge ($Q$), third component of isospin ($I$) \etc, are
carried by different flavours ($u,d,s$) of quarks, in the conventional
quasi-particle models, conserved charges come in strict proportion to
number of $u$, $d$, $s$ quarks. Thus conserved charges are strongly
correlated with the flavours and the flavours have no correlations among
themselves. On the other hand, in the model of \cite{shuryak}, presence
of bound states demand correlations among different flavours. Hence
correlations between conserved charges and flavours depend on the
mass-spectrum of the bound states and the strong correlations among them
are lost. 

Based on the above arguments, in \cite{koch}, it has been suggested that
the quantity
\beq
  C_{BS} = -3 \frac{ \langle BS \rangle - \langle B \rangle \langle S
    \rangle } {  \langle S^2 \rangle - \langle S \rangle^2}, 
\eeq
can be used to probe the degrees of freedom of QGP. Here $B=(U+D-S)/3$
is the net baryon number and  $U$, $D$, $S$ are the numbers of net
(quarks minus anti-quarks) up-quarks, down-quarks and strange-quarks
respectively. The notation $\langle \cdot \rangle$ denotes average taken
over a suitable ensemble. It has been argued in \cite{koch} that for 
QGP where quarks are the effective degrees of freedom,
\ie, where correlations among $U$, $D$ and $S$ are absent, $C_{BS}$ will
have a value of $1$ for all temperature $T>T_c$. On the other hand, for
the model of \cite{shuryak} $C_{BS}=0.62$ at $T=1.5T_c$, while for a gas
of hadron resonances $C_{BS}=0.66$. Thus the knowledge of $C_{BS}$ helps
to identify the degrees of freedom in QGP.

By extending the idea of \cite{koch}, recently in \cite{gavai}, many
ratios like
\beq
  C_{(KL)/L}=\frac{ \langle KL \rangle - \langle K \rangle \langle L
    \rangle } {  \langle L^2 \rangle - \langle L \rangle^2} \equiv
    \frac{ \c_{KL} } { \c_L },
\label{eq.ratio}    
\eeq
have been calculated using lattice QCD simulations with two flavours of
dynamical light quarks and three flavours (two light and one heavy) of
valance quarks. Here $\c_L$ and $\c_{KL}$ denote the susceptibilities
corresponding to conserved charge $L$ and correlation among conserved
charges $K$ and $L$ respectively. The physical meaning of the ratios
like $C_{(KL)/L}$ can be interpreted as follows--- Create an excitation
with quantum number $L$ and then observe the value of a different
quantum number $K$ associated with this excitation. Thus these ratios
identify the quantum numbers corresponding to different excitations and
hence provide information about the degrees of freedom. The calculations
of \cite{gavai} found no evidence for the existence of bound states
\cite{shuryak} even at temperatures very close to $T_c$.  These finding
are consistent with the results of \cite{ejiri}, where the hypothesis of
\cite{shuryak} has been tested by investigating the ratios of higher
order baryon number susceptibilities obtained from lattice simulations.

As these lattice studies \cite{gavai} involved simulations with
dynamical quarks, they were done using small lattices having temporal
lattice size $N_{\tau}=4$. By comparing with the results from quenched
simulations it has been shown \cite{gavai} that $C_{(KL)/L}$ do not
depend on $N_{\tau}$ for temperature $T=2T_c$. It is clearly important
to verify whether the same conclusion holds even close to $T_c$.
Furthermore, it is known that in the case of quenched QCD with standard
staggered quarks the diagonal quark number susceptibilities (QNS) have
strong dependence on the lattice spacing even for the free theory
\cite{gavai1, gavai2}.  On the other hand, the off-diagonal QNS are
identically zero for an ideal gas and acquires non-zero value only in
the presence of interactions. So the lattice spacing dependence of the
off-diagonal QNS is likely to be more complicated, as opposed to that
for the diagonal QNS where these corrections are dominated by the
lattice artifacts of the naive staggered action. Thus if these two QNS
become comparable the ratios mentioned in eq.\ (\ref{eq.ratio}) can
have non-trivial dependence on the lattice spacing $a$ and hence the
continuum limit of these ratios can be different from that obtained
using small lattices. Since the perturbative expressions for diagonal
and off-diagonal QNS (for vanishingly small quark mass and chemical
potential) are respectively \cite{blaizot}---
\beq
  \frac{\c_{ff}}{T^2} \simeq 1 + {\mathcal O}(g^2) 
  \qquad {\rm and} \qquad
  \frac{\c_{ff'}}{T^2} \simeq -\frac{5}{144\pi^6}~ g^6 \ln g^{-1},
\label{eq.pert}
\eeq
it is reasonable to expect that the off-diagonal QNS may not be
negligible at the vicinity of $T_c$ where the coupling $g$ is large. As
the contributions of the bound states in the QNS become more and more
important as one approaches $T_c$ \cite{liao}, on the lattice it is
necessary to investigate the continuum limit of the these ratios of in
order to verify the existence of bound states in a strongly coupled QGP.
At present a continuum extrapolation of this kind can only be performed
using quenched approximation due to the limitations of present day
computational resources. A quenched result for these ratios will also
provide an idea about the dependence of these ratios on the sea quark
mass.

The aim of this work is to carefully investigate the continuum limit of
the ratios of the kind $C_{(KL)/L}$ for temperatures $T_c<T\le2T_c$
using quenched lattice QCD simulations. The plan of this paper is as
follows --- In Section \ref{sc.results} we will give the details of our
simulations and present our results. In the  Section \ref{sc.discussion}
we will summarise and discuss our results.
\section{Simulations and results} \label{sc.results}

The partition function of QCD for $N_f$ flavours, each with chemical
potential $\mu_f$ and mass $m_f$, at temperature $T$ has the form
\beq
  {\cal Z}\left(T,\{\mu_f\},\{m_f\}\right) = \int {\cal DU}~
    e^{-S_G({\cal U})} \prod_f \det M_f (T,\mu_f,m_f), 
\eeq
where $S_G$ is the gauge part of the action and $M$ is the Dirac
operator. We have used standard Wilson action for $S_G$ and staggered
fermions to define $M$. The temperature $T$ and the spatial volume $V$
are expressed in terms of lattice spacing $a$ by the relations
$T=1/(aN_\tau)$ and $V=(aN_s)^3$, $N_s$ and $N_\tau$ being the number of
lattice sites in the spatial and the Euclidean time directions
respectively. The flavour diagonal and the flavour off-diagonal quark
number susceptibilities (QNS) are given by---
\beqa
  \c_{ff} &=& \left(\frac{T}{V}\right) \frac{\partial^2\ln{\cal Z}}
  {\partial \mu_f^2} =  \left(\frac{T}{V}\right) \left[ \left\langle
  {\rm \bf{Tr}} \left( M_f^{-1}M_f'' - M_f^{-1}M_f'M_f^{-1}M_f'\right)
  \right\rangle+\left\langle\left\{ {\rm\bf{Tr}}
  \left(M_f^{-1}M_f'\right) \right\}^2\right\rangle\right], \qquad{\rm
  and} \qquad \\ \c_{ff'} &=& \left(\frac{T}{V}\right)
  \frac{\partial^2\ln{\cal Z}} {\partial \mu_f \partial \mu_{f'}} =
  \left(\frac{T}{V}\right)\left\langle{\rm\bf{Tr}}
  \left(M_f^{-1}M_f'\right){\rm\bf{Tr}}\left(M_{f'}^{-1}M_{f'}'\right)
  \right\rangle ,
\eeqa 
respectively. Here the single and double primes denote first and second
derivatives with respect to the corresponding $\mu_f$ and the angular
bracket denote averages over the gauge configurations.

In this paper we report results of these susceptibilities on lattices
with $N_\tau=4,~8,~10,~{\rm and}~12$, for the temperatures $1.1T_c\le
T\le 2T_c$, chemical potential $\mu_f=0$ and using quenched
approximations. The details of our scale setting procedure are given in
\cite{swagato}. We have generated quenched gauge configurations by using
the Cabbibo-Marinari pseudo-heatbath algorithm with Kennedy-Pendleton
updating of three $SU(2)$ subgroups on each sweep.  Since for
$m_q/T_c\le0.1$ QNS are almost independent of the bare valance quark
mass ($m_q$) \cite{gavai1}, we have used $m_q/T_c=0.1$ for the light $u$
and $d$-flavours. Motivated by the fact that for the full theory
$m_s/T_c\sim1$ we have used $m_q/T_c=1$ for the heavier $s$-flavour. The
fermion matrix inversions were done by using conjugate gradient method
with the stopping criterion $|r_n|^2< \epsilon|r_0|^2$, $r_n$ being the
residual after the $n$-th step and $\epsilon=10^{-4}$ \cite{gupta}. The
traces have been estimated by the stochastic estimator--- ${\rm\bf
Tr}A=\sum_{i=1}^{N_v}R_i^{\dagger}AR_i/2N_v$, where $R_i$ is a complex
vector whose components have been drawn independently from a Gaussian
ensemble with unit variance. The square of a trace has been calculated
by dividing $N_v$ vectors into $L$ non-overlapping sets and then using
the relation--- $({\rm\bf Tr}A)^2=2\sum_{i>j=1}^{L}({\rm\bf
Tr}A)_i({\rm\bf Tr}A)_j/L(L-1)$. We have observed that as one approaches
$T_c$ from above these products, and hence $\c_{ff'}$, become more and
more noisy for larger volumes and smaller quark masses. So in order to
reduce the errors on $\c_{ff'}$ number of vectors $N_v$ have been
increased (for the larger lattices and the smaller quark masses) with
decreasing temperature. Details of all our simulations are provided in
Table \ref{tb.simulation}. 
\begin{table}[!ht]
\squeezetable
\begin{center}
\begin{tabular}{|c|c|c|c|c|c|} \hline

  $T/T_c$&$\beta$&Lattice size&$N_{stat}$&\multicolumn{2}{c|}{$N_v$} \\
  \cline{5-6}
  &&&&$m_q/T_c=0.1$&$m_q/T_c=1$ \\ \hline

  &5.7000&$4\times10^3$&44&250&100 \\
  &&$~\times16^3$&50&250&100\\
  &&$~\times20^3$&30&250&100\\
  1.1&6.1250&$8\times18^3$&48&250&100 \\
  &6.2750&$10\times22^3$&38&250&100 \\
  &6.4200&$12\times26^3$&41&250&100 \\ \hline

  &5.7880&$4\times10^3$&52&100&100 \\
  1.25&6.2100&$8\times18^3$&49&200&100 \\
  &6.3600&$10\times22^3$&46&200&100 \\
  &6.5050&$12\times26^3$&45&200&100 \\ \hline

  &5.8941&$4\times10^3$&51&100&100 \\
  1.5&6.3384&$8\times18^3$&49&150&100 \\
  &6.5250&$10\times22^3$&49&150&100 \\
  &6.6500&$12\times26^3$&48&150&100 \\ \hline

  &6.0625&$4\times10^3$&51&100&100 \\
  2.0&6.5500&$8\times18^3$&50&100&100 \\
  &6.7500&$10\times22^3$&46&100&100 \\
  &6.9000&$12\times26^3$&49&100&100 \\ \hline

\end{tabular}
\end{center}
\caption{The couplings ($\beta$), lattice sizes ($N_\tau\times N_s^3$),
number of independent gauge configurations ($N_{stat}$) and number of
vectors ($N_v$) that have been used for our simulations are given for
each temperature. The gauge configurations were separated by $100$
sweeps.}
\label{tb.simulation}
\end{table}

In the following sections we present our results. The notations we use
are same as in \cite{gavai}. Since we use equal masses for the two light
$u$ and $d$ flavours, the flavour diagonal susceptibilities in this
context are $\c_{uu}=\c_{dd}\equiv\c_u$ and the flavour off-diagonal
susceptibilities are $\c_{du}=\c_{ud}$. For the heavy flavour $s$ the
flavour diagonal susceptibility is denoted as $\c_{ss}\equiv\c_s$ and
the flavour off-diagonal susceptibilities are $\c_{ds}=\c_{us}$.
Expressions for all the susceptibilities used here have been derived in
the appendix of \cite{gavai}.
\subsection{Susceptibilities}
\begin{figure}[h!]
\begin{center}
  \includegraphics[scale=0.7]{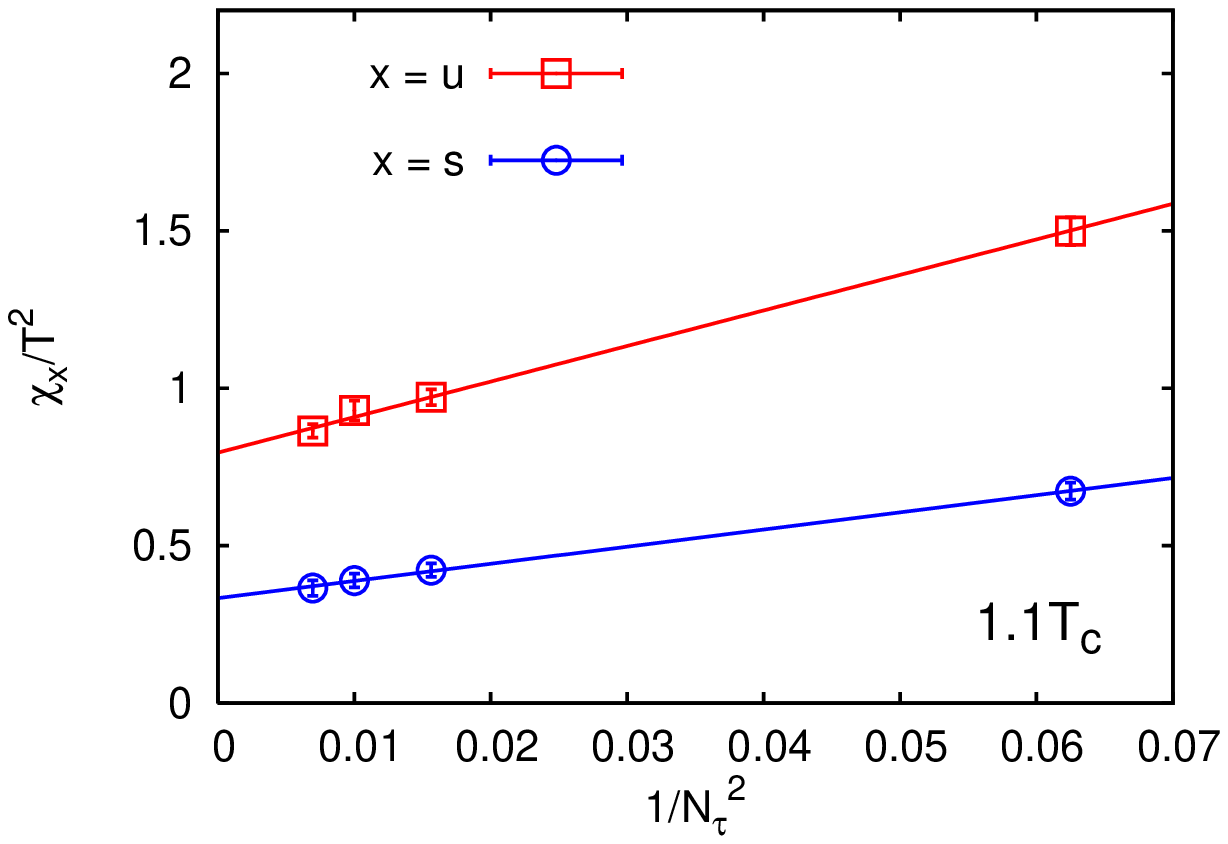}
  \includegraphics[scale=0.7]{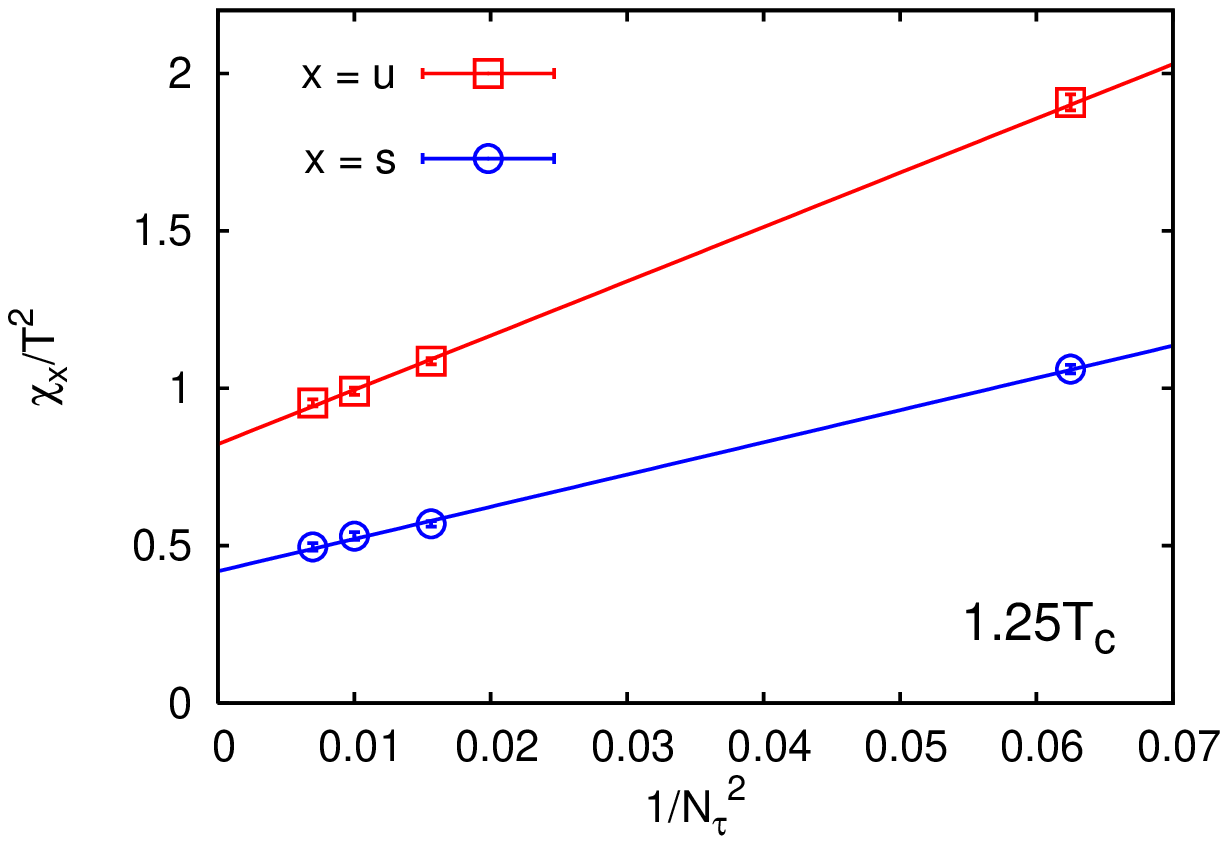}
\end{center}
\caption{We show the $N_\tau$ ($\propto 1/a$) dependence of $\c_u/T^2$
(squares) and $\c_s/T^2$ (circles) for $1.1T_c$ (top panel) and for
$1.25T_c$ (bottom panel).  The continuum extrapolations (linear fits in
$1/N_\tau^2$) are shown by the lines.}
\label{fig.u-s}
\end{figure}
In order to understand the cut-off dependence of $C_{(KL)/L}$ let us
start by examining the same for the diagonal and off-diagonal QNS. We
have found that the for all the temperatures the diagonal QNS ($\c_u$
and $\c_s$) depend linearly on $a^2\propto 1/N_\tau^2$, \ie, the finite
lattice spacing corrections to the diagonal QNS have the form
$\c_{ff}(a,m_f,T)=\c_{ff}(0,m_f,T)+b(m_f,T)a^2+\cdots$. As an
illustration of this we have shown our data for $1.1T_c$ and $1.25T_c$
in Fig.\ \ref{fig.u-s}. Similar variations were found for the other
temperatures also.  We have made continuum extrapolations of the
diagonal QNS by making linear fits in $1/N_\tau^2$. Our continuum
extrapolated results match, within errors, with the available data of
\cite{gavai1} at $1.5T_c$ and $2T_c$.

\begin{figure}[h!]
\begin{center}
  \includegraphics[scale=0.7]{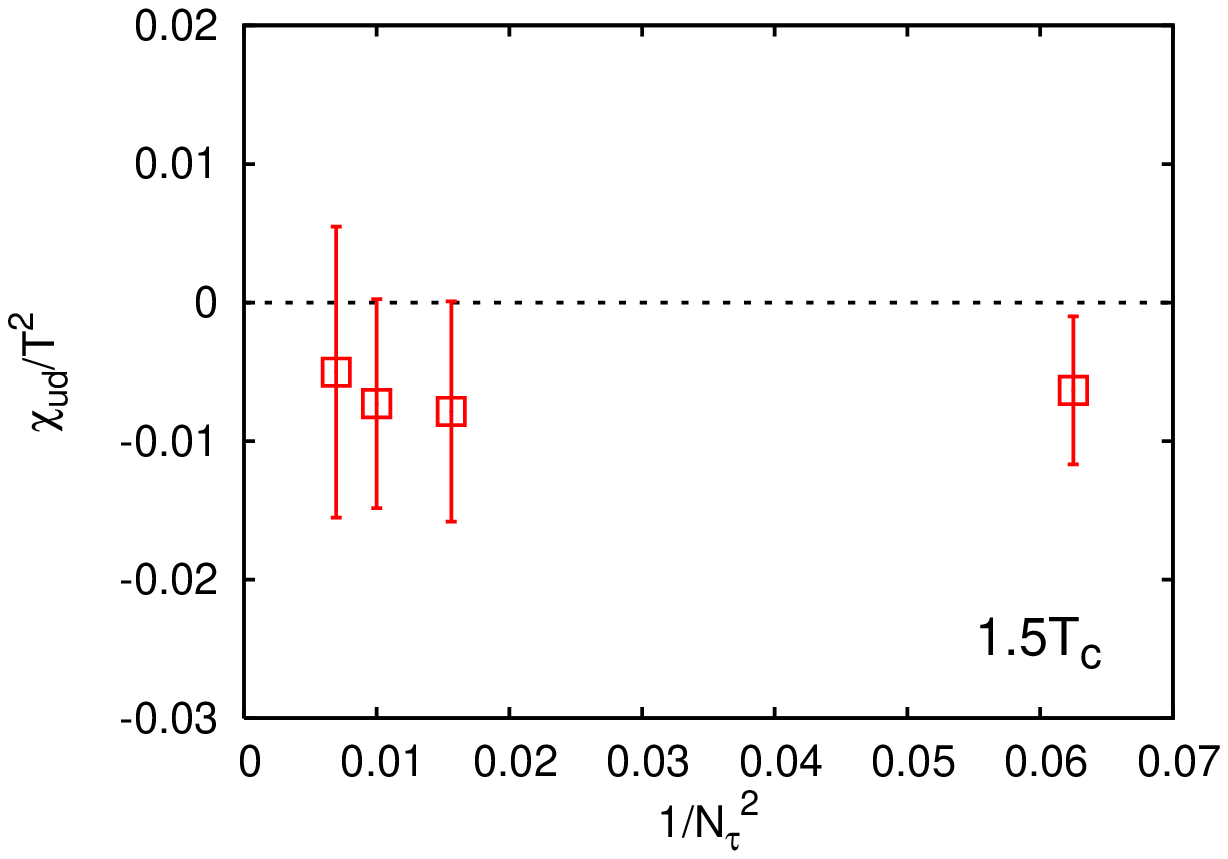}
  \includegraphics[scale=0.7]{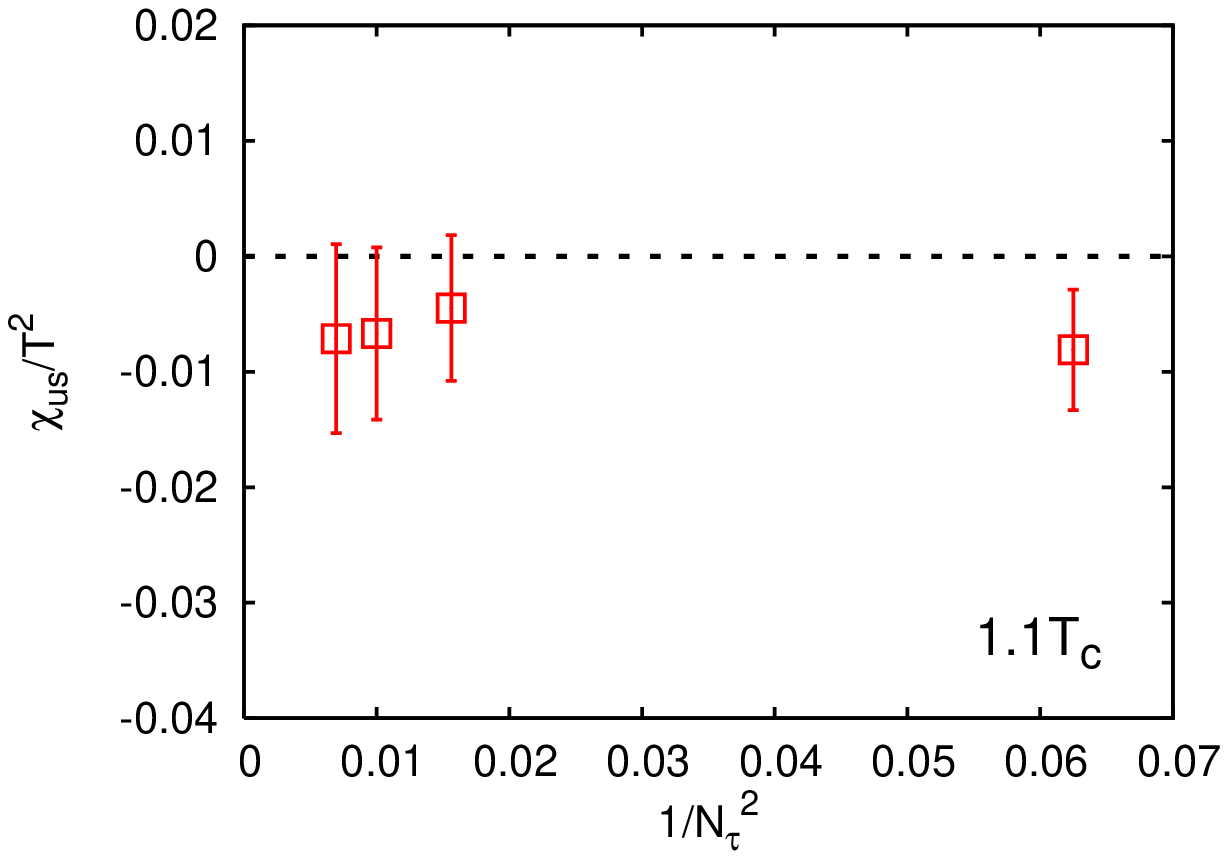}
\end{center}
\caption{$N_\tau$ dependence of the off-diagonal QNS $\c_{ud}/T^2$ at
$1.5T_c$ (top panel) and $\c_{us}/T^2$ at $1.1T_c$ (bottom panel) have
been shown.}
\label{fig.ud-us}
\end{figure}

In Fig.\ \ref{fig.ud-us} we present some of our typical results for the
off-diagonal QNS. Note that here the scales are $\sim100$ magnified as
compared to Fig.\ \ref{fig.u-s}. The sign of our off-diagonal QNS is
consistent with the perturbative predictions of \cite{blaizot}, as well
as with the lattice results of \cite{gupta,ejiri1}. The order of
magnitude of our off-diagonal QNS matches with the results of
\cite{gupta} which uses the same unimproved staggered fermion action as
in the present case.  As can be seen from Fig.\ \ref{fig.ud-us}, within
our errors, we have not found any perceptible dependence $\c_{ff'}$ on
the lattice spacing $a$. Hence to good approximation
$\c_{ff'}(a,m_f,m_{f'},T)\approx\c_{ff'}(0,m_f,m_{f'},T)$. Also for the
other temperatures, which are not shown in Fig.\ \ref{fig.ud-us}, similar
variations were found. Results of our continuum extrapolations of the
diagonal and off-diagonal QNS are listed in Table\ \ref{tb.cont-qns}.
\begin{table}[h!]
\squeezetable
\begin{center}
\begin{tabular}{|c|c c c|c c c|c c|c c|} \hline

  $T/T_c$&\multicolumn{3}{c|}{$\c_u/T^2$}&\multicolumn{3}{c|}{$\c_s/T^2$}&
  \multicolumn{2}{c|}{$\c_{ud}/T^2$}&\multicolumn{2}{c|}{$\c_{us}/T^2$}
  \\ \cline{2-11}
  &$a$&$b$&$\c^2_{d.o.f}$&$a$&$b$&$\c^2_{d.o.f}$& 
  $c\times10^3$&$\c^2_{d.o.f}$&$c\times10^3$&$\c^2_{d.o.f}$ \\ \hline

  1.1&0.79(1)&11.3(5)&0.3&0.33(1)&5.4(1)&0.1&-4(4)&0.5&-6(4)&0.1 \\ [2pt]

  1.25&0.84(1)&15(1)&0.5&0.45(1)&10(1)&0.8&-0.2(1.0)&0.1&-0.7(1.0)&0.6 \\ [2pt]

  1.5&0.83(1)&17.3(3)&0.5&0.55(1)&12.5(5)&0.9&-7(5)&0.1&2(2)&0.6 \\ [2pt]

  2.0&0.86(2)&19.7(2)&0.7&0.70(2)&17(2)&0.8&2(3)&0.5&-0.1(1.0)&0.8 \\ \hline

\end{tabular}
\end{center}
\caption{Parameters for the continuum extrapolations of the diagonal
($\c_u$, $\c_s$) and off-diagonal ($\c_{ud}$, $\c_{us}$) QNS. For the
diagonal QNS continuum extrapolations are made by fitting $a+b/N_\tau^2$
to our data for the three largest lattice sizes. For the off-diagonal
QNS continuum extrapolations are made by fitting our data to a constant
$c$. Numbers in the bracket denote the errors on the fitting parameters
and $\c^2_{d.o.f}$ refers to the value of the chi-square per degrees of
freedom for that particular fit.}
\label{tb.cont-qns}
\end{table}
\begin{figure}[h!]
\begin{center}
  \includegraphics[scale=0.7]{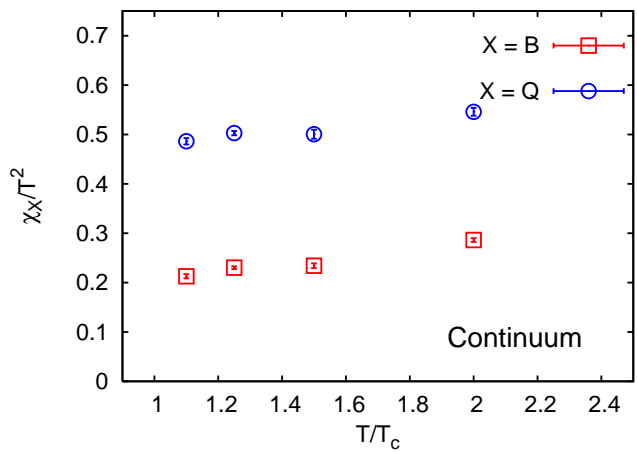}
\end{center}
\caption{The continuum results for $\c_B/T^2$ (squares) and $\c_Q/T^2$
(circles) have been shown.}
\label{fig.b-q}
\end{figure}

For the sake of completeness we also present our continuum extrapolated
results for the two very important quantities, the baryon number
susceptibility ($\c_B$) and the electric charge susceptibility ($\c_Q$).
These quantities are related to the event-by-event fluctuations of
baryon number and electric charge \cite{e-b-e} which have already been
measured at RHIC \cite{e-b-e-expt}. The definitions that we use for
$\c_B$ and $\c_Q$ are \cite{gavai}
\beq
  \c_B = \frac{1}{9}\left( 2\c_u + \c_s + 2\c_{ud} + 4\c_{us} \right) ,
  \qquad{\rm and}\qquad
  \c_Q = \frac{1}{9}\left( 5\c_u + \c_s - 4\c_{ud} - 2\c_{us} \right) .
\eeq
In Fig.\ \ref{fig.b-q} we show the continuum results for $\c_B/T^2$ and
$\c_Q/T^2$. Continuum extrapolations have been performed by making
linear fits in $a^2\propto1/N_\tau^2$. Continuum limit of these
quantities were also obtained in \cite{gavai1} for $T\ge1.5T_c$, though
using different definitions for these quantities. Nevertheless, given
the compatibility of our diagonal QNS with that of \cite{gavai1} and the
smallness of the off-diagonal QNS for $T\ge1.5T_c$ our continuum results
for $\c_B$ and $\c_Q$ are compatible with that of Ref.\ \cite{gavai1}, for any
chosen definitions for these quantities.
\subsection{Ratios}
\begin{figure}[h!]
\begin{center}
  \includegraphics[scale=0.7]{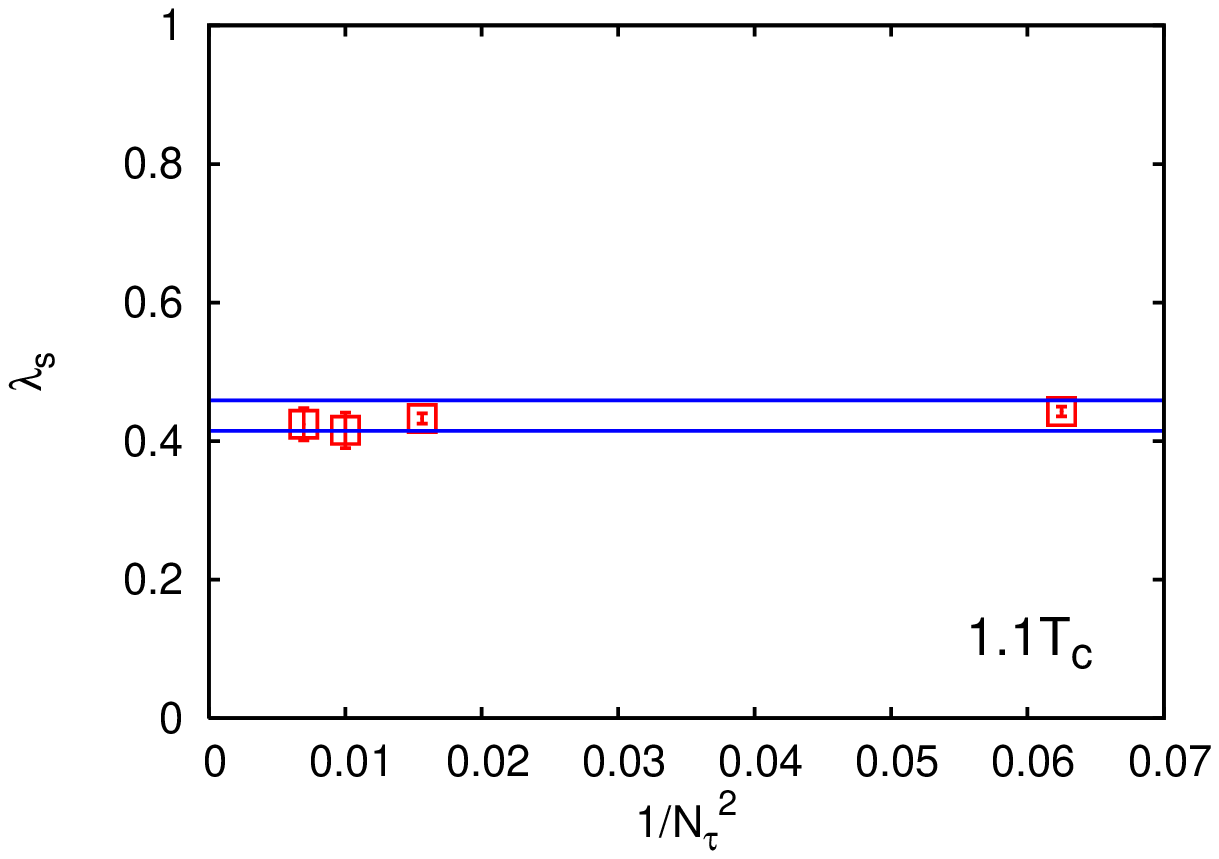}
  \includegraphics[scale=0.7]{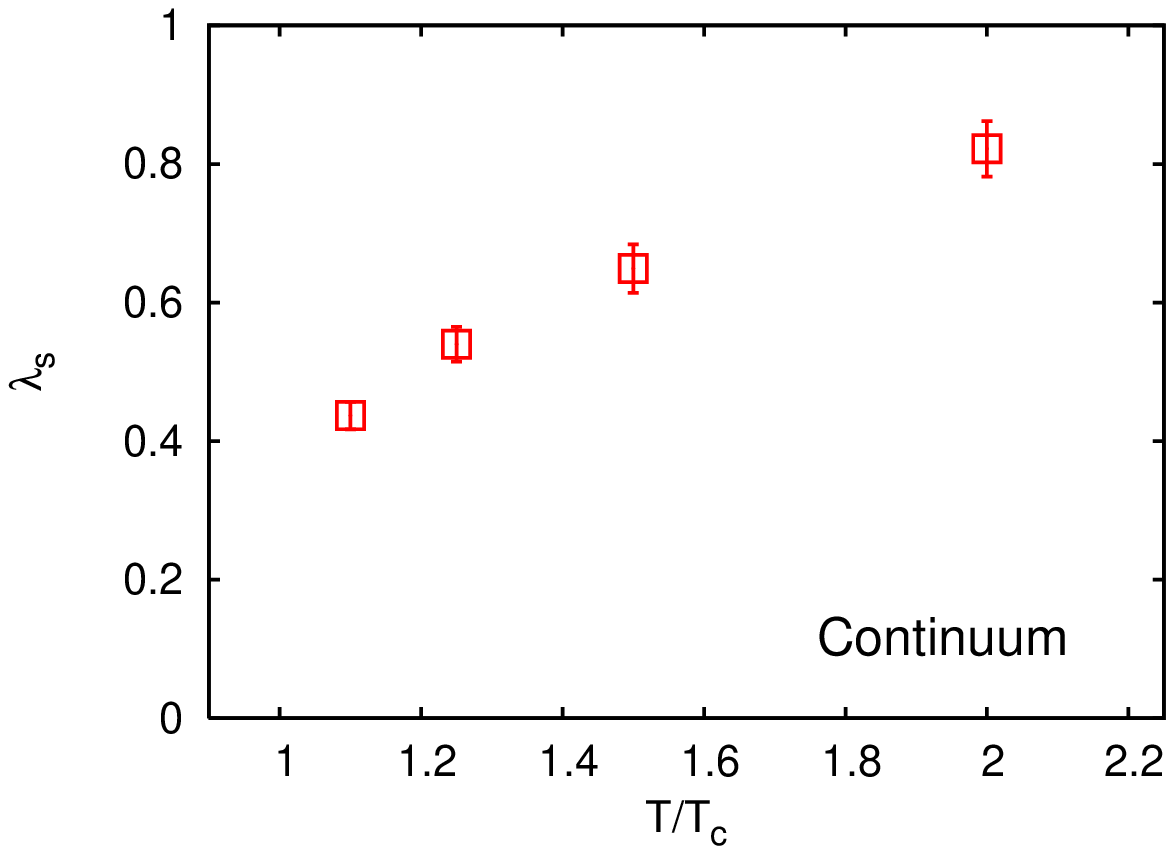}
\end{center}
\caption{In the top panel robustness of the Wroblewski parameter
($\lambda_s$) with changing lattice spacings has been shown for
$1.1T_c$. The lines indicate the $5\%$ error band of a constant fit to
this data. In the bottom panel we show our continuum results for
$\lambda_s$ (see text for details).}
\label{fig.lam_s}
\end{figure}

Wroblewski parameter ($\lambda_s$) \cite{wroblewski} is a quantity of
extreme interest due to its relation to the enhancement of strangeness
production in QGP \cite{rafelski}. The rate of production of quark pairs
in a equilibrated plasma is related to the imaginary part of the
complex QNS by fluctuation-dissipation theorem. If one assumes that the
plasma is in chemical (and thermal) equilibrium and the typical energy
scales for the production of $u$, $d$ and $s$ quarks are well separated
from the inverse of the characteristic time scale of the QCD plasma,
then using Kramers-Kroing relation one can relate $\lambda_s$ to the
ratio of QNS \cite{gavai3}---  
\beq
  \lambda_s = \frac{2\langle s\bar{s} \rangle}
    {\langle u\bar{u} + d\bar{d} \rangle} = \frac{\c_s}{\c_u}.
\eeq
(In the above equation $\langle f\bar{f}\rangle$ should be interpreted
as quark number density and not as quark anti-quark condensates.) We
have found that $\lambda_s$, which is a ratio of two diagonal QNS,
remains constant (within $\sim5\%$) with varying lattice spacings for
all temperatures in $1<T/T_c\le2$. We have illustrated this in the top
panel of Fig.\ \ref{fig.lam_s} by plotting $\lambda_s$ with $1/N_\tau^2$
for the temperature $1.1T_c$.  These results are somewhat surprising
since the order $a^2$ corrections are not negligible for the individual
diagonal QNS. But for the ratio of the diagonal QNS for two different
bare valance quark masses these order $a^2$ corrections happen to be
negligible and thus seems to be quark mass independent. This indicates
that the finite lattice spacing corrections to the diagonal QNS is
constrained to have the form
$\c_{ff}(a,m_f,T)=\c_{ff}(0,m_f,T)[1+b(T)a^2+\cdots]$, as opposed to the
more general form
$\c_{ff}(a,m_f,T)=\c_{ff}(0,m_f,T)+b(m_f,T)a^2+\cdots$.

Our continuum results for the Wroblewski parameter have been shown in
the bottom panel of Fig.\ \ref{fig.lam_s}. In view of the constancy of
$\lambda_s$ we have made the continuum extrapolations by making a
constant fit to $a^2\propto1/N_\tau^2$. Our Continuum limit for
$\lambda_s$ are consistent with the previously reported \cite{gavai1}
continuum values for $T\ge1.5T_c$.  Our continuum results for
$\lambda_s$ are very close to the results of \cite{gavai} for the whole
temperature range of $T_c<T\le2T_c$. Closeness of our quenched results
with the results from the dynamical simulations of \cite{gavai} suggest
that the Wroblewski parameter has practically no dependence on the mass
of the sea quarks.  These observations along with the fact that
$\lambda_s$ has very mild dependence on the valance quark mass
\cite{gupta1} shows that the present day lattice QCD results for the
Wroblewski parameter are very reliable. The robustness of the Wroblewski
parameter is very encouraging specially since in the vicinity of $T_c$
the lattice results for this quantity almost coincides with the value
($\lambda_s \approx 0.43$) extracted by fitting the experimental data of
RHIC with a hadron gas fireball model \cite{cleymans}.
\begin{figure}[h!]
\begin{center}
  \includegraphics[scale=0.7]{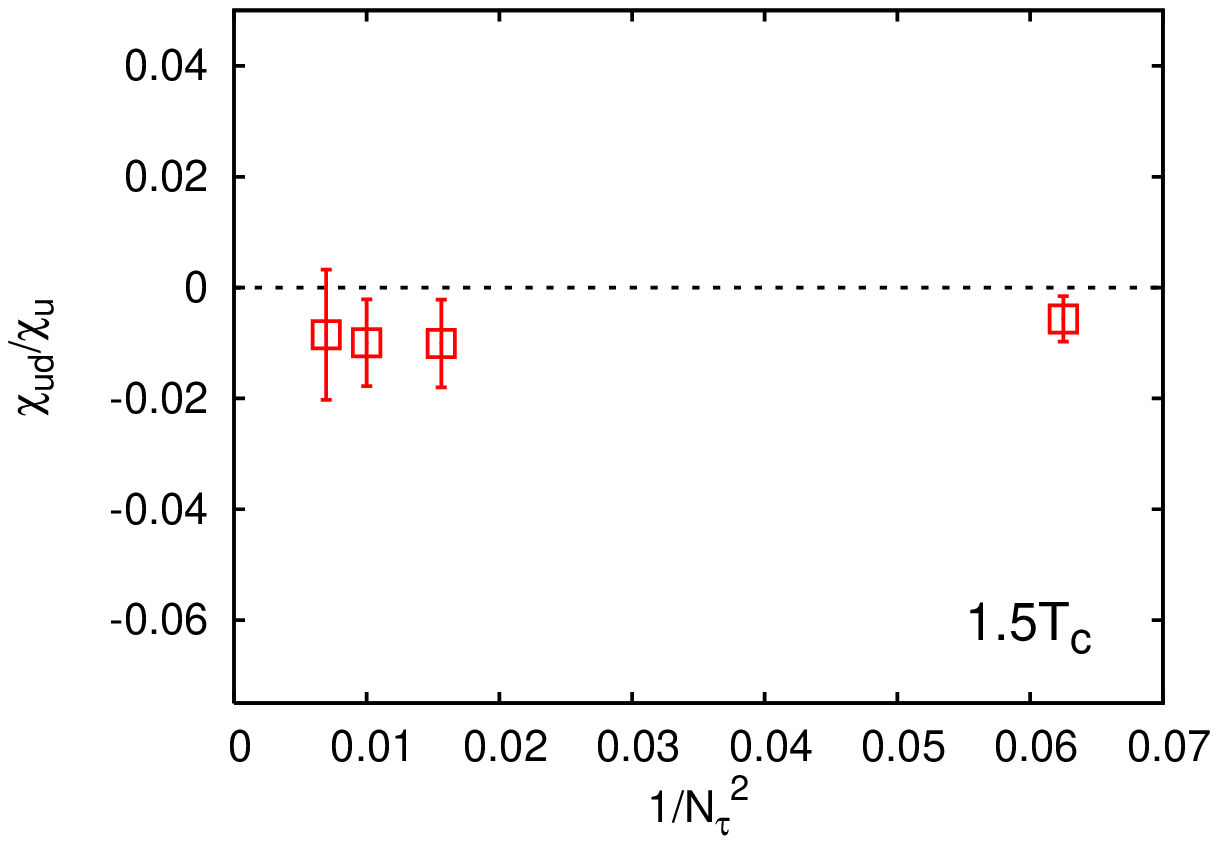}
  \includegraphics[scale=0.7]{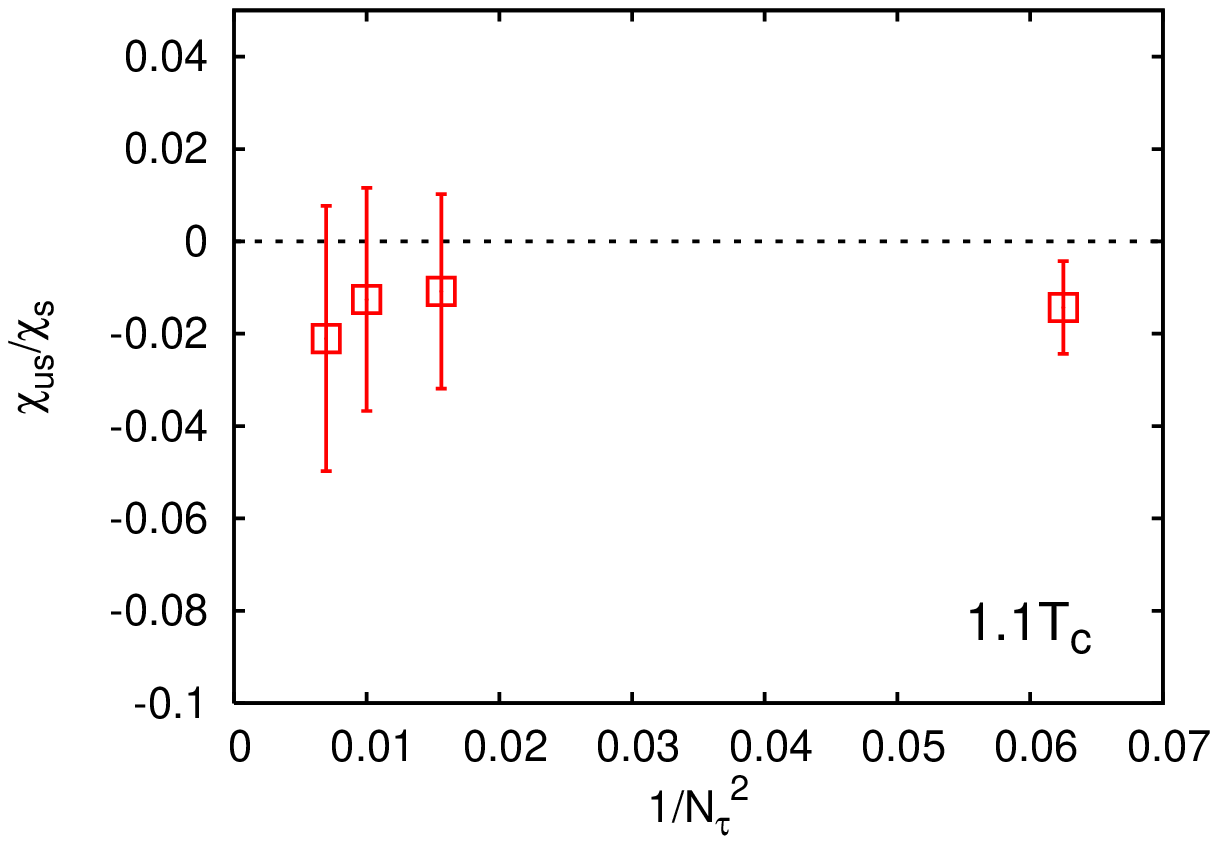}
\end{center}
\caption{The top panel shows the $N_\tau$ dependence of $\c_{ud}/\c_u$
at $1.5T_c$. The bottom panel shows the same for $\c_{us}/\c_s$ at
$1.1T_c$ }
\label{fig.ud_u-us_s}
\end{figure}

After examining the ratio of the diagonal QNS let us focus our attention
on the ratios of off-diagonal to diagonal QNS.  Given our results for
the diagonal and off-diagonal QNS it is clear that these will have the
form--- $\c_{ff'}(a,m_f,m_{f'},T)/\c_{ff}(a,m_f,T)\approx
[\c_{ff'}(0,m_f,m_{f'},T)/\c_{ff}(0,m_f,T)][1-b(T)a^2]$. Since $b(T)$ is
positive, \ie, $\c_{ff}$ decreases with decreasing lattice spacing, this
ratio is expected to decrease (as $\c_{ff'}$ is negative) and move
away from zero. However, due smallness of these ratios itself, within
our numerical accuracies, we have been unable to identify any such
effect.  This has been exemplified in Fig.\ \ref{fig.ud_u-us_s} where
$\c_{ud}/\c_u$ at $1.5T_c$ (top panel) and  $\c_{us}/\c_s$ at $1.1T_c$
(bottom panel) have been shown.
\begin{figure}[t!]
\begin{center}
  \subfigure[]{ \includegraphics[scale=0.68]{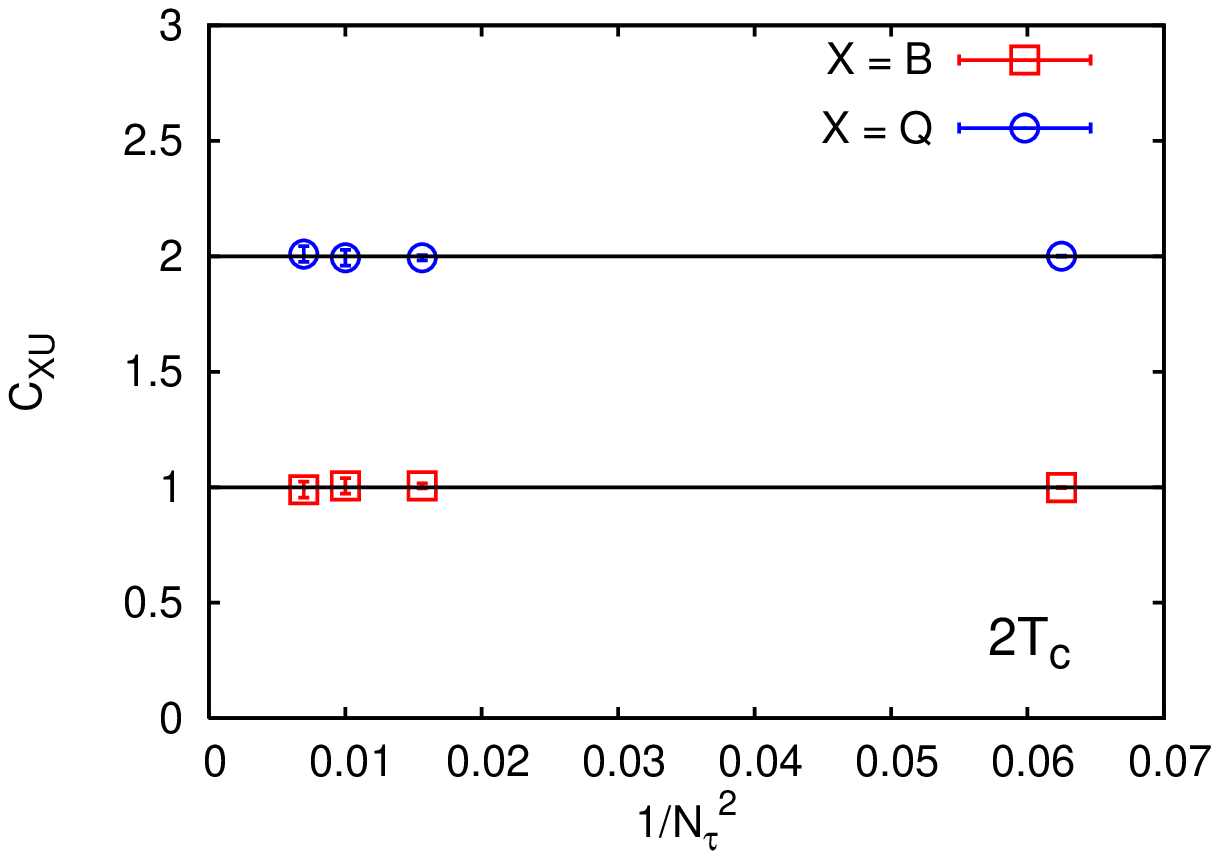} }
  \subfigure[]{ \includegraphics[scale=0.68]{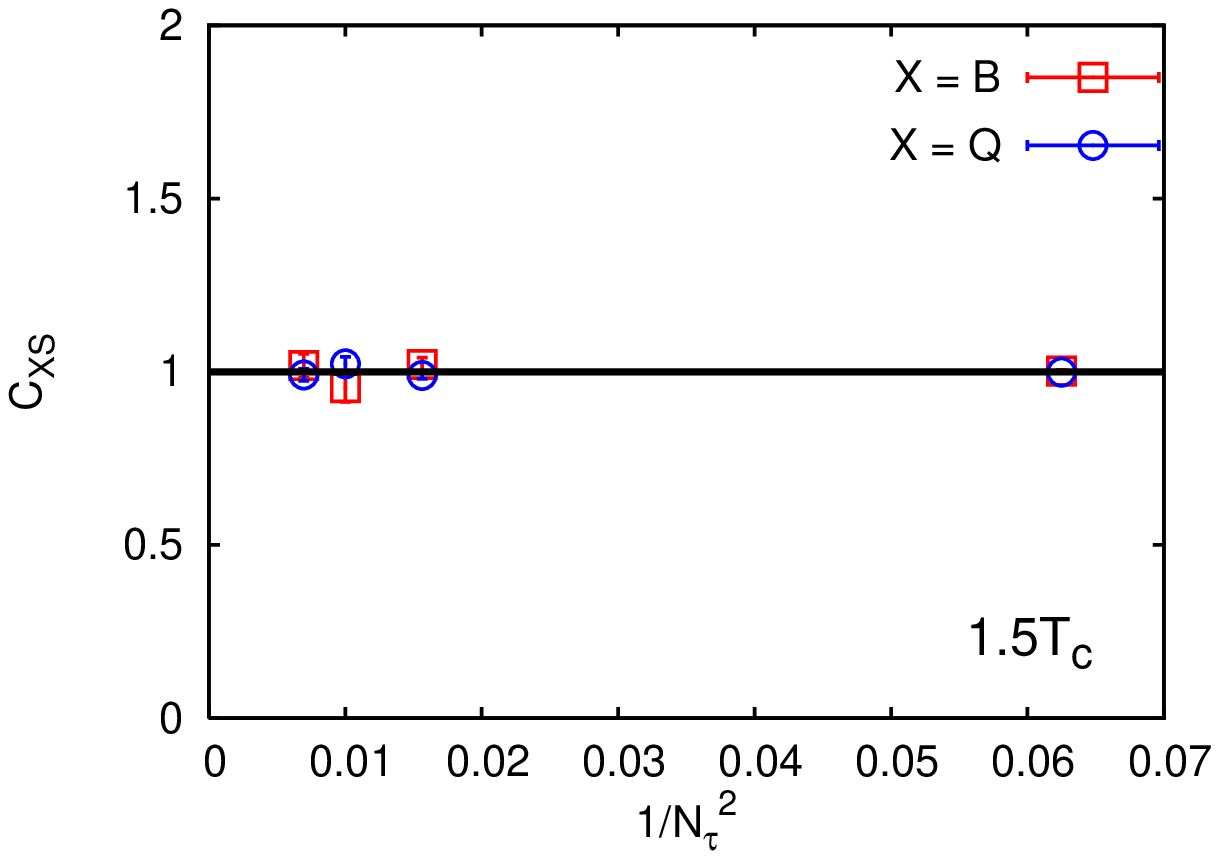} } \\
  \subfigure[]{ \includegraphics[scale=0.68]{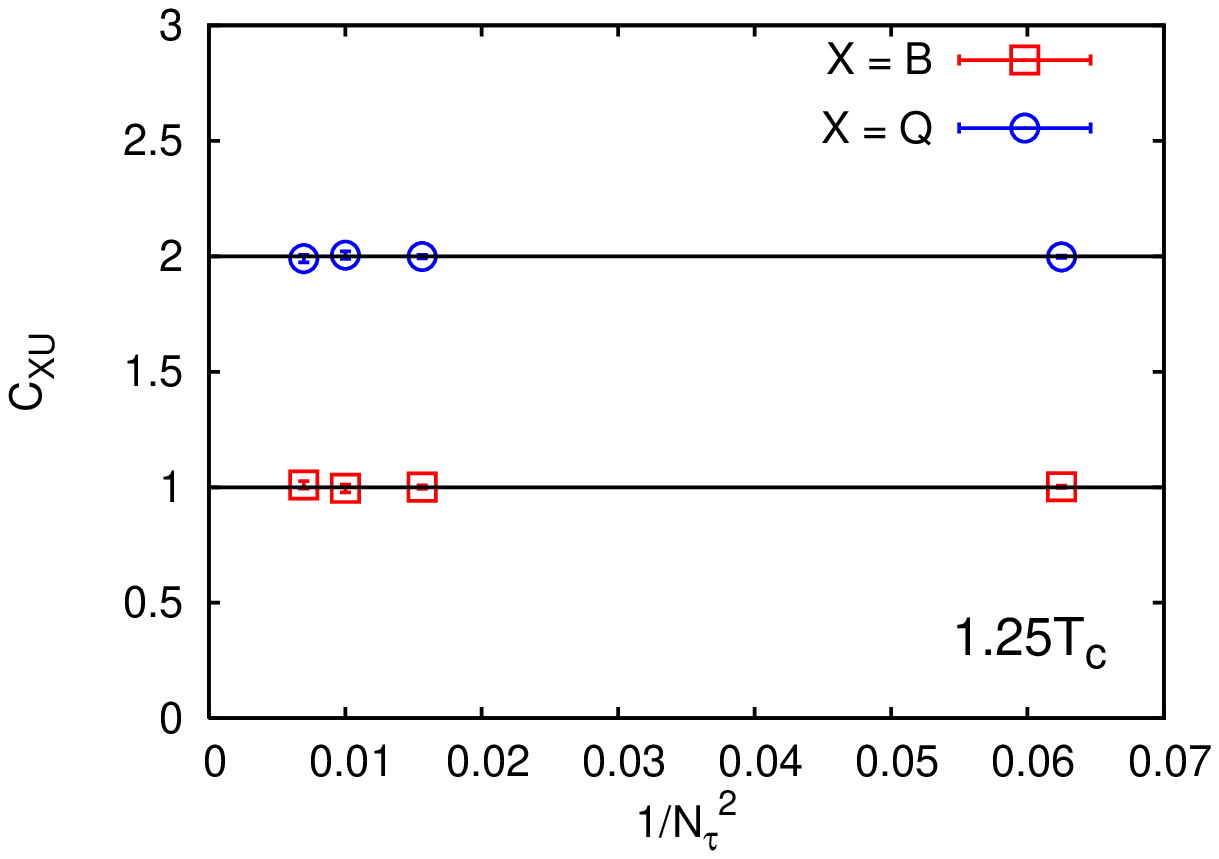} }
  \subfigure[]{ \includegraphics[scale=0.68]{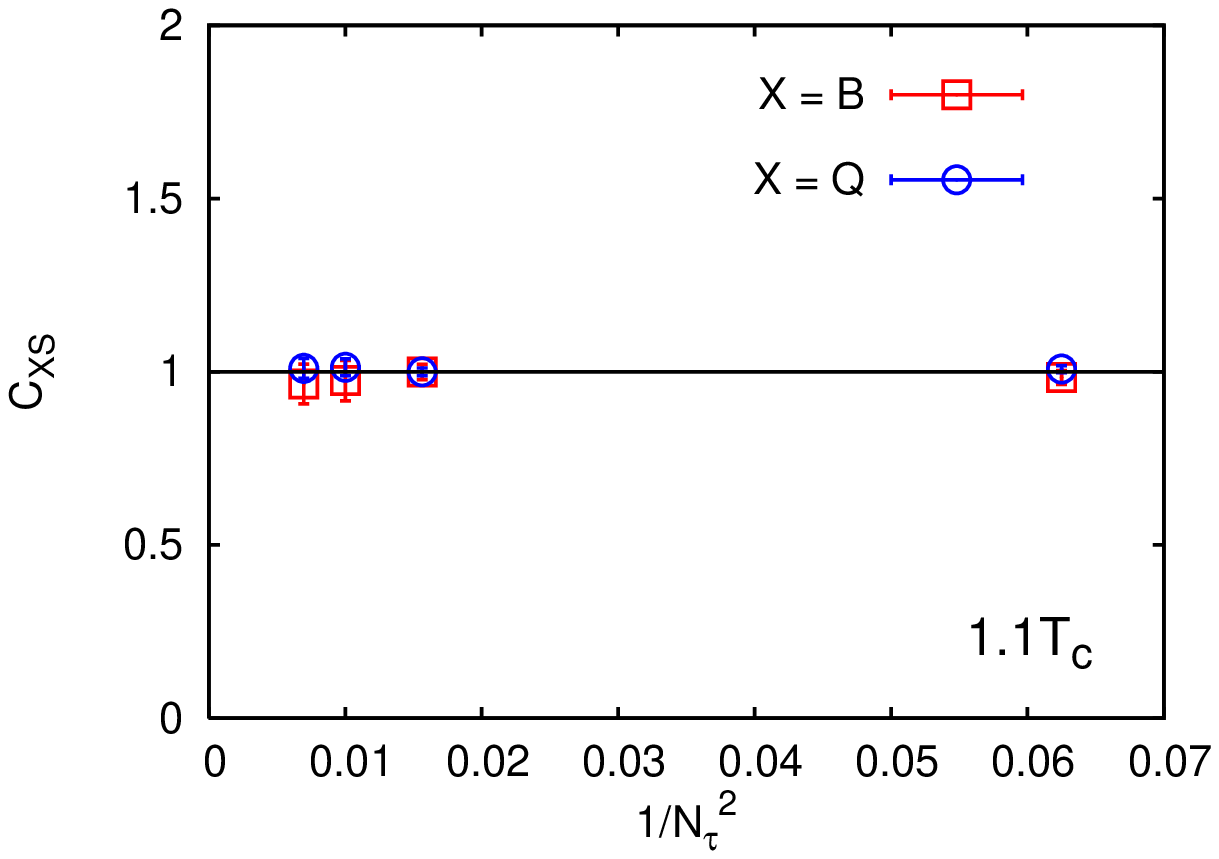} }
\end{center}
\caption{Lattice spacing dependence of $C_{XU}$ and $C_{XS}$ are shown
for temperatures $2T_c$ [panel (a)], $1.5T_c$ [panel (b)], $1.25T_c$
[panel (c)] and $1.1T_c$ [panel (d)] by plotting these quantities as a
function of $1/N_\tau^2$ ($\propto a^2$), for $N_\tau=4,8,10,12$. The
lines indicate the ideal gas values for these ratios.}
\label{fig.cxy}
\end{figure}

Following the main theme of this paper we now present the lattice
spacing dependence of ratios the like $C_{(KL)/L}$. Two such ratios that
can directly probe the degrees of freedom in a QGP are \cite{koch, gavai}
\begin{subequations}
\beqa
  C_{BS} &\equiv& -3C_{(BS)/S} = -3\frac{\c_{BS}}{\c_S} = 
   \frac{\c_s +2\c_{us}}{\c_s} = 1 + \frac{2\c_{us}}{\c_s} ,
  \qquad{\rm and}\qquad \\
  C_{QS} &\equiv& 3C_{(QS)/S} = 3\frac{\c_{QS}}{\c_S} = 
   \frac{\c_s - \c_{us}}{\c_s} = 1 - \frac{\c_{us}}{\c_s}  .
\eeqa
\label{eq.cxs}   
\end{subequations}
These quantities probe the linkages of the strangeness carrying
excitations to baryon number ($C_{BS}$) and electric charge ($C_{QS}$)
and hence give an idea about the average baryon number and the average
electric charge of all the excitations carrying the $s$ flavours.  These
ratios are normalized such that for a pure quark gas, \ie, where unit
strangeness is carried by excitations having $B=-1/3$ and $Q=1/3$,
$C_{BS}=C_{QS}=1$. A value of $C_{BS}$ and $C_{QS}$ significantly
different from $1$ will indicate that the QGP phase may contain
some other degrees of freedom apart form the quasi-quarks.

Similar ratios can also be formed for the light quark sector
\cite{gavai}, \eg, for the $u$ flavour the ratios
\begin{subequations}
\beqa
 C_{BU} &\equiv& 3 C_{(BU)/U} = 3\frac{\c_{BU}}{\c_U} = 
   \frac{ \c_u + \c_{ud} + \c_{us} }{\c_u} = 
   1 + \frac{\c_{ud}}{\c_u} + \frac{\c_{us}}{\c_u},
 \qquad{\rm and}\qquad  \\
 C_{QU} &\equiv& 3 C_{(QU)/U} = \frac{3\c_{QU}}{\c_U} =
  \frac{ 2\c_u - \c_{ud} - \c_{us} } {\c_u} = 
  2 - \frac{\c_{ud}}{\c_u} - \frac{\c_{us}}{\c_u}
\eeqa
\label{eq.cxu}    
\end{subequations}
quantifies the average baryon number ($C_{BU}$) and and the average
electric charge ($C_{QU}$) of all the excitations carrying $u$ quarks.
For a medium of pure quarks, \ie, where the $u$ flavours are carried by
excitations with baryon number $1/3$ and electric charge $2/3$,
$C_{BU}=1$ and $C_{QU}=2$. Similar ratios can also be formed for the
$d$ quarks \cite{gavai}.
As can be seen seen from eqs.\ (\ref{eq.cxs}, \ref{eq.cxu}) the lattice

spacing dependence of $C_{BS}$ \etc\ are governed by the cut-off
dependence of the ratios $\c_{ff'}/\c_{ff}$. Since we have already
emphasised that, within our numerical accuracies, the ratios
$\c_{ff'}/\c_{ff}$ are almost independent of lattice spacings it is
expected that the same will also happen for the ratios $C_{(KL)/L}$. In
accordance to this expectation we have found that for temperatures
$1.1T_c\le T\le 2T_c$ these ratios are independent of lattice spacings
within $\sim 5\%$ errors, see Fig.\ \ref{fig.cxy}. Note that these
ratios are not only independent of the lattice spacings but also acquire
values which are very close to their respective ideal gas limits.
\begin{figure}[t!]
\begin{center}
  \includegraphics[scale=0.7]{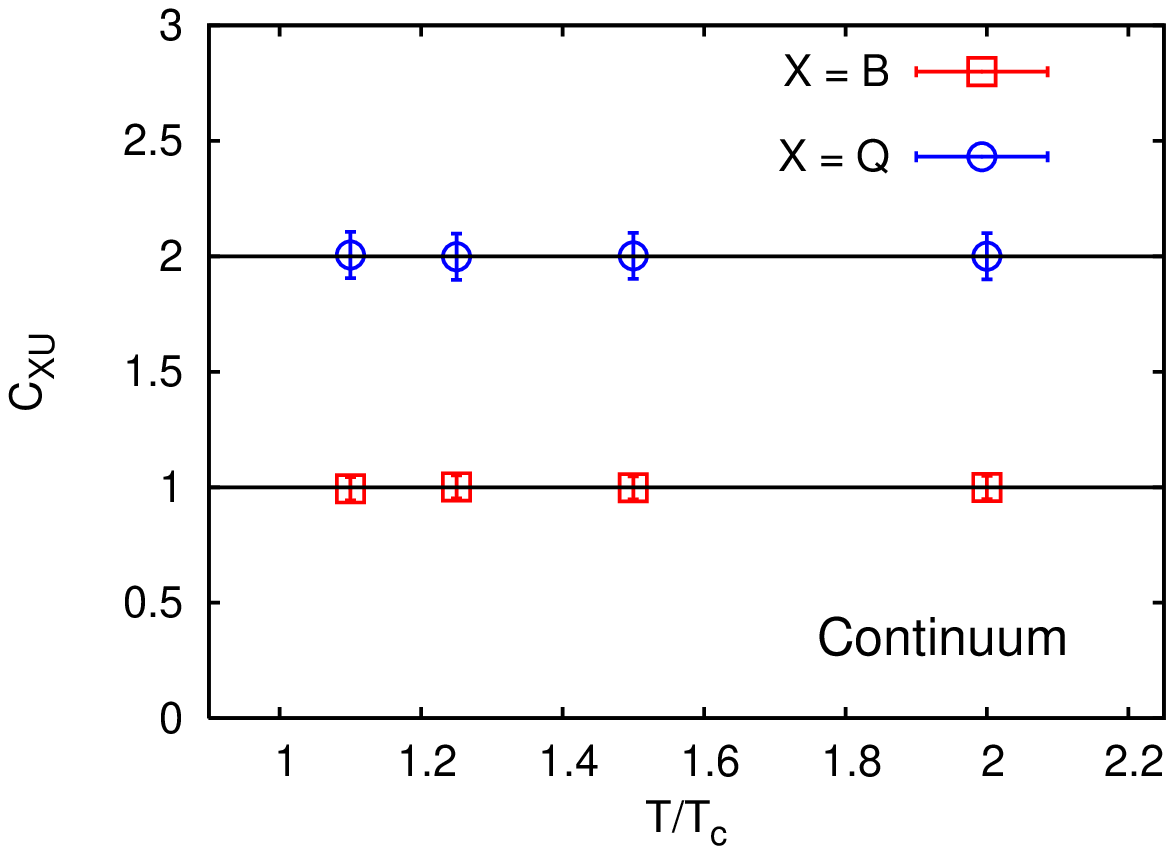}
  \includegraphics[scale=0.7]{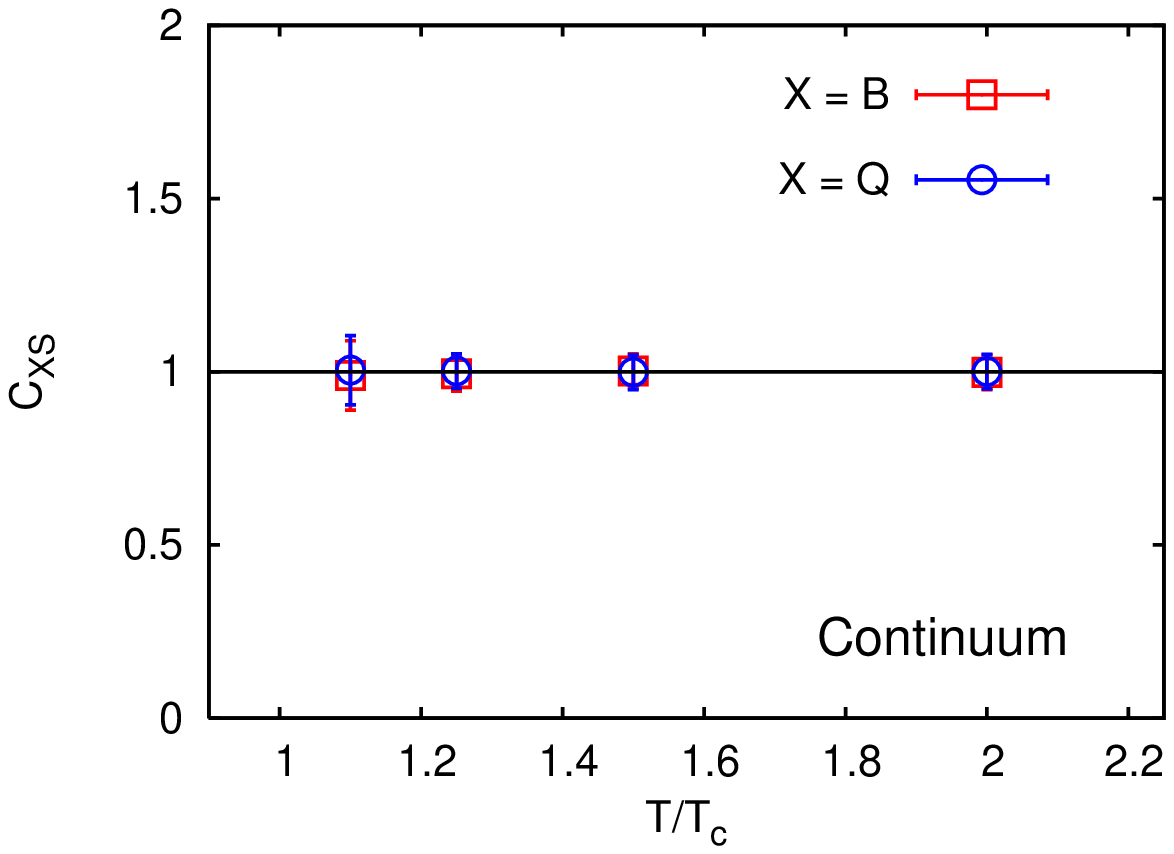}
\end{center}
\caption{Continuum results for $C_{XU}$ (top panel) and $C_{XS}$ (bottom
panel). The lines indicate the ideal gas values for these quantities.
See text for details.}
\label{fig.cxy-cont}
\end{figure}

In Fig.\ \ref{fig.cxy-cont} we present our continuum results for
$C_{XS}$ (bottom panel) and $C_{XU}$ (top panel), where $X=B,Q$. Since
these ratios remain almost constant with changing $1/N_\tau^2$ (see
Fig.\ \ref{fig.cxy}) we have made continuum extrapolations by making
constant fits of our data to $1/N_\tau^2$. For the whole temperature
range of interests ($T_c<T\le 2T_c$) these ratios have values which are
compatible with that for a gas of pure quarks. This is exactly what has
been found in \cite{gavai} using partially quenched simulations with
smaller lattices. For the $d$ quarks also we have found similar results.
\section{Summary and discussion} \label{sc.discussion}
In this paper we have made a careful investigation of the continuum
limit of different ratios of off-diagonal to diagonal susceptibilities
in quenched QCD using lattices with large temporal extents
($N_\tau=12,10,8~ {\rm and}~ 4$), for a very interesting range of
temperature ($T_c<T\le2T_c$) and for vanishing chemical potential. We have
found that for this whole range of temperature the lattice results for
the ratios like $C_{BS}$, $C_{QS}$ \etc\ are robust, \ie, they are
almost independent (within $\sim5\%$) of the lattice spacing.  We have
also arrived at the same conclusion for the Wroblewski parameter which
is of interest to the experiments in RHIC and Large Hadron Collider
(LHC).

At this point, it is good to have some idea about how unquenching may
change our results. It has been found \cite{gavai4} that in the
temperature range $T\ge1.25T_c$ there is only $5-10\%$ change in the QNS
in going from quenched to $N_f=2$ dynamical QCD. On the other hand,
since the order of the phase transition depends strongly on the number
of dynamical flavours the change in QNS is likely to be much larger in
the vicinity of the transition temperature for the quenched theory which
has a first order phase transition. Though this may be true for the
individual QNS, their ratios may have very mild dependence on the sea
quarks content of the theory. Given the good compatibility of our
results of $C_{BS}$, $C_{QS}$ \etc\ with the results of \cite{gavai} it
is clear that indeed these ratios have very mild dependence on the sea
quark content of the theory. It is also known that \cite{gavai1} for
bare valance quark mass of $m_q/T_c\le0.1$ the dependence of the QNS on
the valance quarks mass is very small. Hence our results show that the
ratios like $C_{(KL)/L}$ are robust not only in the sense that they do
not depend on the lattice spacings but also they have very mild
dependence on the quark masses.
\begin{figure}[t!]
\begin{center}
  \includegraphics[scale=0.7]{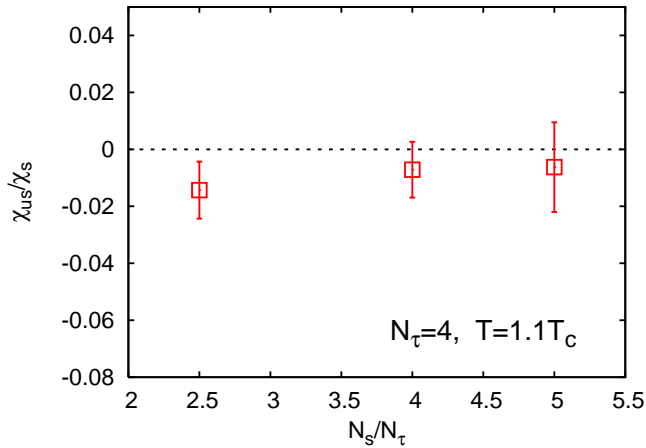}
\end{center}
\caption{Dependence of the ratio $\c_{us}/\c_s$ on the aspect ratio has
been shown for $N_\tau=4$ at temperature $1.1T_c$.}
\label{fig.vol-dep}
\end{figure}

All the results presented in this paper are for spatial lattice sizes
$N_s=2N_\tau+2$, \ie, for aspect ratios $N_s/N_\tau=2.5-2.17$. In view
of the fact that quenched QCD has a first order phase transition it is
important to have some idea about the volume dependence of our
results, specially in the vicinity of the transition temperature $T_c$.
To check this dependence we have performed simulations using lattices
having aspect ratios $N_s/N_\tau=2.5-5$, for our smallest temporal
lattice $N_\tau=4$ and at temperature $1.1T_c$. In these simulations we
have not found any significant volume dependence of any quantity which
have been presented in this paper. As an illustration, in Fig.\
\ref{fig.vol-dep}, we have shown the dependence of $\c_{us}/\c_s$ on the
aspect ratio, for $N_\tau=4$ at $1.1T_c$. The volume dependence is 
expected to be even smaller as one goes further away from first order
phase transition point. Also the agreement of our results with that of
\cite{gavai}, where an aspect ratio of $4$ have been used, shows that
the these ratios have almost no volume dependence for $N_s\ge N_\tau+2$.  

While the closeness of $C_{XU}$ and $C_{XS}$ ($X=B,Q$) to their
respective ideal gas values do support the notion of quasi-particle like
excitations in QGP, a significant deviation of these ratios from their
ideal gas values neither rule out the quasi-particle picture nor
confirms the existence of the bound states proposed in \cite{shuryak}.
Large contributions from the chemical potential dependence of the
quasi-particle masses may lead to significant deviation of these ratios,
especially in the vicinity of $T_c$. It has already been pointed out
\cite{bluhm,liao} that, near $T_c$, the chemical potential dependence of
the quasi-particle masses becomes crucial for the baryonic
susceptibilities.

Nevertheless, it may be interesting to compare our results with the
predictions of the bound state model of \cite{shuryak}. Based on the
model of \cite{shuryak} (and assuming that the mass formulae given in
\cite{shuryak} hold right down to $T_c$) the predicted values of
$C_{BS}$ are approximately $0.62$ at $1.5T_c$ \cite{koch}, $0.11$ at
$1.25T_c$ and almost zero at $1.1T_c$ \cite{ack-majumder}. Clearly, as
can be seen form Fig.\ \ref{fig.cxy-cont} (bottom panel), these values
are very much different from our continuum results. However, it has been
argued in \cite{liao} that apart from all the bound states mentioned in
\cite{shuryak}, baryon like bound states may also exist in QGP. These
baryons make large contributions to the baryonic susceptibilities,
especially close to $T_c$ \cite{liao}. Taking account the contributions
from the strange baryons may increase the value of $C_{BS}$. In
\cite{liao} it has also been argued that for two light flavours if one
considers the contributions of the baryons only then close to $T_c$ the
ratio of 2-nd order isospin susceptibility ($d_2^I$) to the 2-nd order
baryonic susceptibility ($d_2$) is
$d^I_2/d_2=(\c_u-\c_{ud})/(\c_u+\c_{ud})=0.467$.  Clearly this is
inconsistent with our results since a value of $d^I_2/d_2=0.467$ gives a
positive $\c_{ud}/\c_u$ ($=0.363$). Whereas, the lattice results for
$\c_{ud}/\c_u$ are negative and much smaller in magnitudes. This suggest
that the contribution of the mesons (also possibly of the quarks,
diquarks and $qg$-states) are definitely important in the isospin
susceptibility $d^I_2$. If one takes into account of the contributions
of the mesons (pions and rhos) and assumes that the Boltzmann weight of
the mesons are equal to that of the baryon one gets a lower bound for
$d^I_2/d_2$, namely $d^I_2/d_2 \ge 0.644$ \cite{ack-liao}. But this
lower bound gives $\c_{ud}/\c_u=0.217$ and hence very far from our
results. Moreover, very recently it has been argued \cite{majumder} that
one can carefully tune the densities of the baryon and meson like bound
states in the model of Refs.\ \cite{shuryak,liao} to reproduce the
lattice results for off-diagonal QNS. But even those carefully tuned
values fail to reproduce \cite{majumder} the lattice results for higher
order susceptibilities. In view of all these, the lattice results of
\cite{gavai} favours a quasi-particle like picture of QGP, as opposed to
the bound state model of \cite{shuryak,liao}. The results of this paper
show that these lattice results are really robust in the sense that they
have very mild dependence on the lattice spacing and sea quark content
of the theory. 
\begin{acknowledgments}
The author is grateful to Rajiv Gavai for his constant encouragement,
many illuminating discussion and careful reading of the manuscript. The
author would also like to thank Sourendu Gupta for many useful comments
and discussion. Part of this work was done during a visit to ECT*,
Trento. The financial support from the Doctoral Training
Programme of ECT*, Trento is gratefully acknowledged.
\end{acknowledgments}
\end{document}